\documentclass[prb,showpacs,twocolumn,amsmath,amssymb,floatfix]{revtex4}

\usepackage{graphicx} \usepackage{dcolumn} \usepackage{bm}

\begin{document}

\title{Image resonance in the many-body density of states at a metal
surface}

\author{G. Fratesi}
\altaffiliation{Present address: SISSA/ISAS, via Beirut 4,
34014 Trieste, Italy.}
\author{G. P. Brivio}
\affiliation{INFM and Dipartimento di Scienza dei Materiali,\\
Universit\`a di Milano-Bicocca, via Cozzi 53, 20125 Milano, Italy}

\author{Patrick Rinke}
\altaffiliation{Present address: Fritz-Haber-Institut der
Max-Planck-Gesellschaft, Faradayweg 4--6, 14195 Berlin-Dahlem, Germany}
\author{R. W. Godby}
\affiliation{Department of Physics, University of York,
York YO10 5DD, United Kingdom}

\date{\today}

\begin{abstract}
The electronic properties of a semi-infinite metal surface without a
bulk gap are studied by a formalism able to account for the continuous
spectrum of the system.
The density of states at the surface is calculated within the $GW$
approximation of many-body perturbation theory.
We demonstrate the presence of an unoccupied surface resonance peaked
at the position of the first image state.
The resonance encompasses the whole Rydberg series of image states and
cannot be resolved into individual peaks.
Its origin is the shift in spectral weight when many-body 
correlation effects are taken into account.
\end{abstract}


\maketitle

%
%
\section{introduction}

For the understanding of several fundamental
properties of condensed matter surfaces, the knowledge of the electronic
density of states and of the parallel-wavevector-resolved density are
essential.
The dispersion of both occupied and excited surface
states, the influence of the changes of the electronic structure in
thin film growth and enhanced magnetism are just a few examples of
relevant quantities.
Experimentally the density of states can be probed by a variety of 
techniques such as
photoemission spectroscopies\cite{mat98} (angle resolved,
integrated, inverse and two-photon), and scanning tunneling
microscopy.\cite{sch01}
To provide an adequate theoretical description of the experimental observables,
it is necessary to employ methods which retain the continuous
character of the spectrum.
In other words one has to take into account that certain quantities may
be defined for any energy in a given energy interval.
Furthermore, the formalism must be capable of describing excited-state
properties.

Regarding excited states, considerable experimental interest has been
devoted to image-potential induced (IPI)
states\cite{dos84,str84,gie85,her96}
and resonances.\cite{str86,qui93,yan93,pet99}
IPI states are present in systems where a bulk band gap provides a
barrier, trapping electrons in the image tail of the surface potential.
If no gap is present at the IPI energies, an electron is not reflected
completely at the bulk barrier and hybridization with surface truncated 
bulk states becomes possible.
This results in the formation of resonances for some materials. 
A comprehensive theoretical description of IPI states has been given
by Echenique and coworkers,\cite{ech78,ech85,bau86,chu98} whereas the
situation for IPI resonances is much less satisfactory.
In principle, IPI resonances are intrinsically contained in the many-body framework,
already at the level of the $GW$ approximation,\cite{hed69} but to
our knowledge surfaces have only been investigated in this context
using a repeated slab geometry.\cite{egu92,dei93,hei98}
Such a simplified treatment cannot capture the continuous spectrum of a
real surface, because the spectral function will inevitably be composed of
a limited number of sharp, discrete peaks in place of the resonance.
%

%
%
%

In this Article we present {\em ab initio} many-body calculations of
the local density of states (LDOS) of a semi-infinite jellium surface
and demonstrate the presence of a broad IPI resonance.
We calculate the LDOS -- in its many-body generalization, the spectral
function\cite{hed69} -- decomposed according to the surface parallel
wave-vector $\mathbf{k}_{\parallel}$ within the surface $xy$-plane as
\begin{equation}\label{eq:ldosA}
    \sigma(\mathbf{r},\omega)=\int\frac{\mathrm{d}^2
    \mathbf{k}_\parallel}{(2\pi)^2} A(z,\mathbf{k}_\parallel,\omega),
\end{equation}
where $A$ is the $\mathbf{k}_{\parallel}$-resolved LDOS or spectral
weight function, defined by
\begin{equation}\label{eq:A}
    A(z,\mathbf{k}_\parallel,\omega)=-\frac{1}{\pi}
    \Im{}G(z,z,\mathbf{k}_\parallel,\omega) \textrm{sgn}(\omega-\mu).
\end{equation}
Here $G$ is the one-particle Green's function in the representation
indicated, and Hartree atomic units ($a_0=0.529$~\AA, 1~hartree$=27.2$~eV)
are used.

$G$ is obtained from Dyson's equation
\begin{equation}\label{eq:dyson}
    G=G^{\textrm{DFT}}+G^{\textrm{DFT}}\big[
    \Sigma_{\textrm{XC}}-v_{\textrm{XC}}\big]G,
\end{equation}
where $v_{\textrm{XC}}$ is the exchange and correlation DFT potential
and $\Sigma_{\textrm{XC}}$ is the $GW$ electron self-energy.

Equation~(\ref{eq:dyson}) is solved using a recent method developed to
perform $GW$ calculations in infinite, non-periodic geometries\cite{fra03}
based on the embedding method.\cite{ing81}
The advantage of this approach is that the semi-infinite substrate,
surface and vacuum regions are treated equally without the need for
any fitting parameters or a repeated cell geometry.

The main steps of the computation are as follows:
(i) The Kohn-Sham equation\cite{koh65} is solved self-consistently
within LDA.\cite{per81}
In this framework, the embedding method permits an exact treatment
of the semi-infinite substrate.
$G^{\textrm{DFT}}$, obtained by numerical inversion of the Kohn-Sham
Hamiltonian, describes a continuous spectrum of fictitious non-interacting
electrons.
(ii) From $G^{\textrm{DFT}}$ we compute the polarization $P$ in the
random phase approximation.
(iii) The inverse dielectric response, $(1-vP)^{-1}$, yields the
effective interaction $W^{\textrm{RPA}}$.
(iv) The self-energy
$\Sigma_{\textrm{XC}}=iG^{\textrm{DFT}}W^{\textrm{RPA}}$ is calculated
using a real-space, imaginary-frequency representation and one obtains
the self-energy on the real frequency axis by means of analytic
continuation.
(v) Equation~(\ref{eq:dyson}) can now be solved to update $G$.

In Sec.~\ref{sec:results} we present and discuss the results.
Section~\ref{sec:conclusions} is devoted to conclusions.
%

%
%
%
\section{\label{sec:results}Results and discussion}

\begin{figure}
\includegraphics[width=1.0\columnwidth]{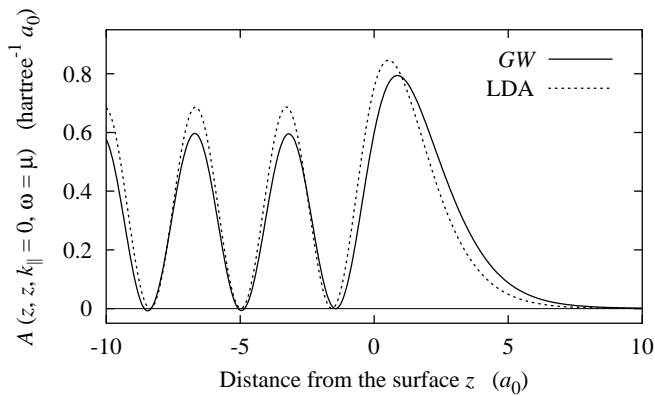}
\caption{\label{fig1:imgzz} Spectral weight function at the chemical
potential and zero parallel wave-vector for semi-infinite jellium with
$r_s=2.07$~$a_0$.}
\end{figure}

We examine first the $z$-dependence of the spectral weight function
$A(z,\mathbf{k}_{\parallel},\omega)$.
This quantity is proportional to the probability amplitude for a
particular wave-vector $\mathbf{k}_{\parallel}$ and energy $\omega$.
In Fig.~\ref{fig1:imgzz} we report
$A(z,\mathbf{k}_{\parallel},\omega)$ for $k_{\parallel}$ equal to zero
and $\omega$ equal to the chemical potential $\mu$.
All results shown in this Article are obtained for a jellium substrate
of aluminum density ($r_s=2.07$~$a_0$).
The spectral weight outside the surface is enhanced by the improved
description of  exchange-correlation effects in $GW$, in common with
the states of Al(111) studied in Ref.~\onlinecite{whi98}.
By varying the energy $\omega$ we have verified that this feature is
common to all bound states.
In the bulk, the $GW$ spectral weight is lower than that 
calculated by
the LDA because some weight is transferred to lower energies through
electron-plasmon coupling.
The larger amplitude of $A$ in the surface layer, together with the
absence of decay into the bulk, identifies  the states in this part of
the surface band structure as forming a surface resonance.
%

\begin{figure}
\includegraphics[width=1.0\columnwidth]{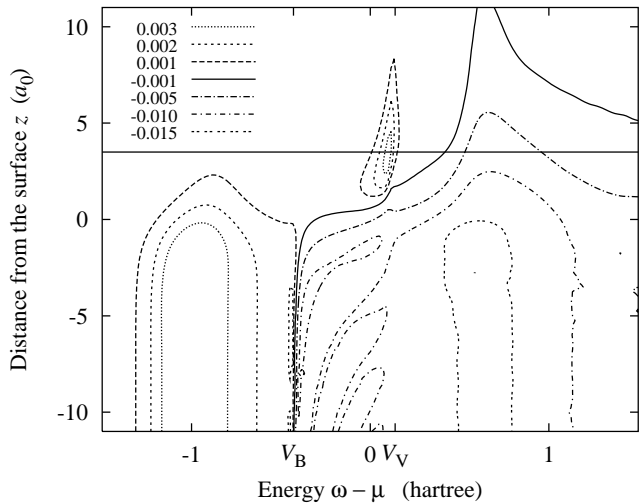}
\caption{\label{fig2:ldos} Contour levels of $\Delta\sigma(z,\omega)$
for semi-infinite jellium with $r_s=2.07$~$a_0$. The solid horizontal
line indicates $z=3.5$~$a_0$ (see Fig.~\ref{fig4:ldos_3.5}).  The
pronounced ``teardrop'' feature at $z \sim 3.5$ $a_0$ is the surface
image resonance. $V_\textrm{V}$ is the vacuum energy level, and
$V_\textrm{B}$ the bottom
of the free-electron band.}
\end{figure}
\begin{figure}
\includegraphics[width=1.0\columnwidth]{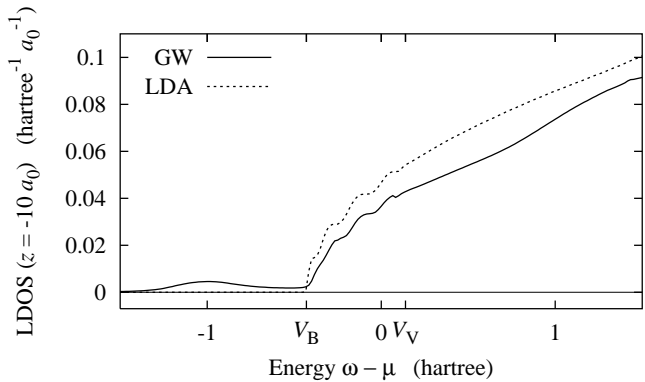}
\caption{\label{fig3:ldos_-10} LDOS $\sigma^{GW}$ and
$\sigma^{\textrm{LDA}}$ for semi-infinite jellium with
$r_s=2.07$~$a_0$ at $z=-10$~$a_0$ from the jellium edge.
The ``ripples'' between $V_\textrm{V}$ and $V_\textrm{B}$
indicate the presence of Friedel oscillations.}
\end{figure}

In Fig.~\ref{fig2:ldos} we plot
$\Delta\sigma=\sigma^{GW}-\sigma^{\textrm{LDA}}$, the difference
between our many-body LDOS and that in the LDA.
The energy-dependence of $\Delta\sigma$ as $z$ moves from the bulk
towards the vacuum is indicated by contour levels.
We recapitulate that, for a jellium surface, the LDA effective
potential $v_{\textrm{eff}}(z)$ approaches the constant limits $V_\textrm{B}$
and $V_\textrm{V}$ for $z\rightarrow-\infty$ (bulk) and $z\rightarrow+\infty$
(vacuum) respectively.
As a consequence, the bulk and vacuum limits of
$\sigma^{\textrm{LDA}}$ are simply proportional to the well-known
expressions $\sqrt{\omega-V_\textrm{B}}$ and $\sqrt{\omega-V_\textrm{V}}$.
In the bulk, the LDA therefore predicts no states below $V_\textrm{B}$.
The electron-plasmon coupling, automatically included in $GW$, 
is responsible for moving
states down in energy from above to below $V_\textrm{B}$, yielding the sub-band
shown in Fig.~\ref{fig3:ldos_-10} for energies around 1~hartree below the
chemical potential, as already demonstrated for the homogeneous
electron gas.\cite{hed69}
%


The presence of the surface is still noticeable even some atomic units
into the substrate through Friedel oscillations, which are visible for
bound energies from the contour levels of Fig.~\ref{fig2:ldos} and in
the LDOS at a distance from the surface $z=-10$~$a_0$ in
Fig.~\ref{fig3:ldos_-10}.
%


When $z$ approaches the surface, coupling with plasmons considerably
reduces, eventually becoming negligible a few atomic units outside the
surface.
As $z$ moves into the vacuum, the electron density decays to
zero. Self-energy effects are no longer present and the LDA
description becomes exact. We then obtain $\Delta\sigma=0$.
%

\begin{figure}
\includegraphics[width=1.0\columnwidth]{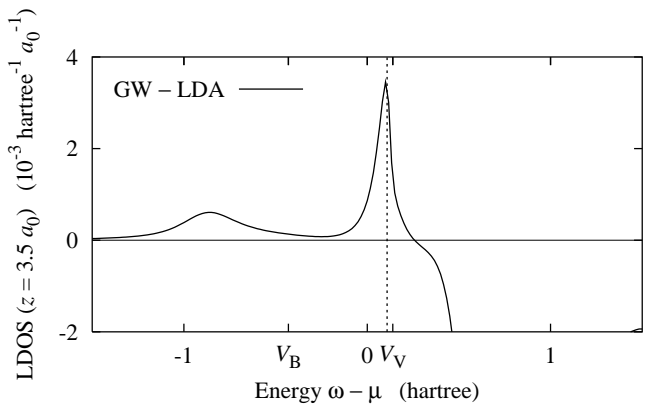}
\caption{\label{fig4:ldos_3.5} Difference in the LDOS $\Delta\sigma$
for semi-infinite jellium with $r_s=2.07$~$a_0$ at $z=3.5$~$a_0$ from
the surface. The dashed vertical line indicates the position of the
first Rydberg state, $1/32$~hartree below the vacuum level.}
\end{figure}

Apart from surface-truncated bulk structures, Fig.~\ref{fig2:ldos}
also highlights a completely new feature in the form of a teardrop-shaped 
enhancement of the LDOS, localized in the near-surface region at energies
between the chemical potential and the vacuum level.
To highlight this feature we plot the energy dependence of $\Delta\sigma$ for 
$z=3.5$~$a_0$ in Fig.~\ref{fig4:ldos_3.5} (note that this corresponds to a
cross section plot along a horizontal line through Fig.~\ref{fig2:ldos}).
The peak in $\Delta\sigma$, located at about $1/32$~hartree below the
vacuum level (the position of the first Rydberg state\cite{ech78}),
clearly identifies an IPI surface resonance.
%


The presence of IPI resonances, even in the limiting case of a
substrate without bulk reflectivity (the bulk reflectivity is associated with
scattering by the atomic-cores\cite{pap90}), has been unclear.
Simple model potentials in independent-particle approximations
produced IPI resonances for jellium substrates\cite{lin89} in some
cases, but the presence of a clear resonance depended sensitively on
the precise details of the chosen model.
We remind that IPI states and resonances are a many-body phenomenon
and can thus be included only in an {\em ad hoc} way in
single-particle dynamic approximations.
In contrast to these earlier, parameter-dependent results, our method
is fully {\em ab initio} and includes many-body correlations.
%


In the case of semi-infinite jellium, the zero bulk reflectivity
yields a large linewidth of the resonance, since hybridization with
bulk states is not prevented at all.
We estimate of the full width at half maximum (FWHM) from $\Delta\sigma$
as $0.07$~hartree (about $2$~eV).
\footnote{Experimentally, the FWHMs of surface resonances at metal surfaces are
usually of the order of tenths of eV.\cite{qui93,bul94}
The difference may be explained by the non-negligible reflectivity of the
atomic cores, reducing the hybridization with the underlying bulk continuum
of states and thus the FWHM.}
We remark that in systems without a bulk band gap the hybridization with the
continuous bulk band is the major contributor to the linewidth, as the
self-energy alone accounts only for about one tenth of the total
FWHM.\cite{chu98}

\begin{figure}
\includegraphics[width=1.0\columnwidth]{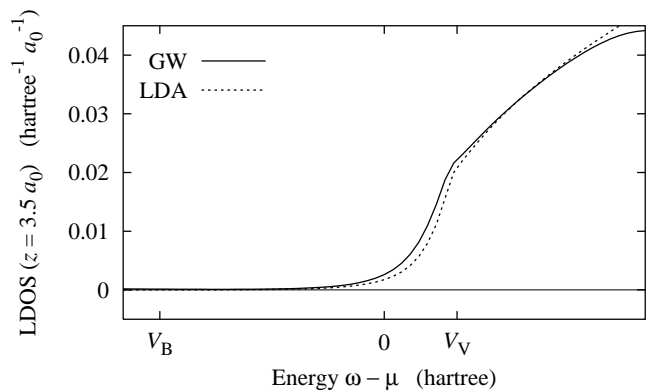}
\caption{\label{fig5:totldos} Local density of states for a
semi-infinite jellium with $r_s=2.07$~$a_0$ at $z=3.5$~$a_0$.
The quantity reported in Fig.~\ref{fig4:ldos_3.5} is the difference
between the continuous and the dashed line.}
\end{figure}

Owing to its large linewidth, the resonance shown in Fig.~\ref{fig4:ldos_3.5}
encompasses the whole series of
Rydberg image states,\cite{qui93} which in the infinite bulk barrier model
extends from $V_\textrm{V}-1/32$~hartree to $V_\textrm{V}$.
The resonance is clearly visible in the difference plot between the many-body 
LDOS and the LDOS in the LDA. In the many-body LDOS itself, on the other hand,
it resolves as a shoulder feature just below the vacuum energy
(Fig.~\ref{fig5:totldos}) rather than a distinct peak, because the LDA
LDOS is already a strongly increasing function of energy in this range.
The quantity reported in Fig.~\ref{fig4:ldos_3.5} is the difference between
two theoretical quantities and is therefore not directly accessible to
experiment.
However, it allows us to
isolate an important contribution to the phenomenon -- the many-body
interaction -- which could
improve the description of the electronic structure of real
semi-infinite surfaces beyond the usual single-particle approaches such as
Ref.~\onlinecite{ish01}.

\begin{figure}
\includegraphics[width=1.0\columnwidth]{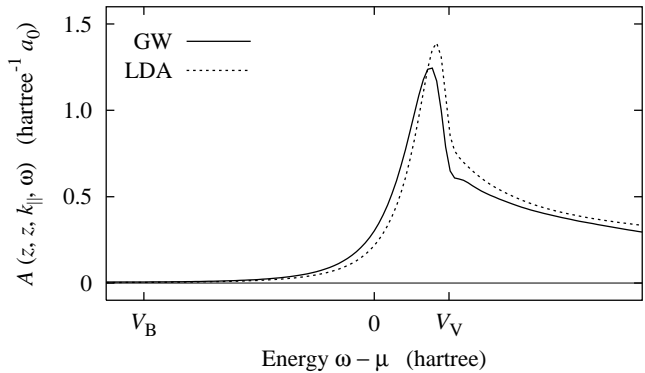}
\caption{\label{fig6:kldos} Spectral weight function for a
semi-infinite jellium with $r_s=2.07$~$a_0$, for $k_\parallel=0$ and
$z=3.5$~$a_0$. The integral over $\mathbf{k}_\parallel$ of this function
gives the LDOS reported in Fig.~\ref{fig5:totldos}.}
\end{figure}

We will now examine the origin of the resonance emerging in
Fig.~\ref{fig4:ldos_3.5}.
For this reason it is important to look at the spectral weight
function $A$, whose integral over $\mathbf{k}_\parallel$ is the LDOS
[Eq.~(\ref{eq:ldosA})].
In Fig.~\ref{fig6:kldos} we present the energy dependence of
$A(z,\mathbf{k}_\parallel,\omega)$ for $\mathbf{k}_\parallel=0$ at the
same position outside the jellium edge as Figs.~\ref{fig4:ldos_3.5}
and~\ref{fig5:totldos} ($z=3.5$~$a_0$).

We first consider the LDA result.
For non-interacting particles in a constant potential $V$ the
$z$-resolved spectral weight function is essentially one-dimensional
and hence proportional to $(\omega-k_{\parallel}^{2}/2-V)^{-1/2}$.
Thus in the vacuum a singularity is present at $V_\textrm{V}$  for
$k_\parallel=0$.
The spectral weight then diminishes with increasing energy.
But near the surface weight is transferred to bulk states decaying out
into the vacuum: the spectral function has a peak very close to
$v_{\textrm{eff}}(z)$, but singularities are no longer present.
We emphasize that in a slab geometry Fig.~\ref{fig6:kldos} would
appear as a collection of delta functions, each delta corresponding to
one of the discrete eigenstates.

When we describe the system in the interacting picture ($GW$), we
still observe bulk states that spill out into the vacuum.
But we also have a new class of states, constituting the IPI resonance.
They have substantial weight at and outside the surface.
As a consequence spectral weight from the LDA states is transferred
into these quasiparticle states.
Spectral weight hence moves down from higher energies to the energies
of the IPI resonance.
This produces the shoulder in $A$ at $V_\textrm{V}$, and the displacement of
the peak energy to lower energies.

If $k_{\parallel}$ is increased from its zero value, the profile of
$A$ shown in Fig.~\ref{fig6:kldos} is shifted towards higher energies
by $k_{\parallel}^{2}/2$, without being distorted too much (i.e., the
dispersion of the states is nearly parabolic with effective mass equal
to $1$).
When we integrate over $k_\parallel$ to get the LDOS, one notices that
for energies roughly below $V_\textrm{V}$ the $GW$ spectral weight $A^{GW}$ is
greater than $A^{\textrm{LDA}}$ for any value of $k_\parallel$,
leading to the positive $\Delta\sigma$ shown in
Fig.~\ref{fig4:ldos_3.5}.
For higher energies, $A^{GW} - A^{\textrm{LDA}}$ changes from negative
to positive as $k_{\parallel}$ increases.
Negative contributions dominate and the corresponding $\Delta\sigma$
is negative.
%

%
%
%
\section{\label{sec:conclusions}Conclusions}

In summary, we have evaluated from first principles the local density
of states of a semi-infinite simple metal surface.
Many-body correlations are included via the $GW$ self-energy.
The substrate is described as truly semi-infinite, thus enabling a
calculation of the continuous spectrum necessary to properly account
for the hybridization between surface electronic states and bulk
states.
We demonstrate the presence of an IPI resonance just below the vacuum
energy, encompassing the full series of image-states that would be present in a
system with a surface bandgap.
The origin of the resonance is explained in terms of the spectral weight
tranfer to lower energies due to the inclusion of electron-electron correlation.


\begin{acknowledgments}
We are grateful to S. Modesti, L. G. Molinari, G. Onida, and
M. I. Trioni for useful discussions. This work was supported by the
Italian MIUR through Grant No. 2001021128. Patrick Rinke acknowledges the
support of EPSRC and the DAAD.
\end{acknowledgments}

\bibliography{imagedos}

\begin{thebibliography}{27}
\expandafter\ifx\csname natexlab\endcsname\relax\def\natexlab#1{#1}\fi
\expandafter\ifx\csname bibnamefont\endcsname\relax
  \def\bibnamefont#1{#1}\fi
\expandafter\ifx\csname bibfnamefont\endcsname\relax
  \def\bibfnamefont#1{#1}\fi
\expandafter\ifx\csname citenamefont\endcsname\relax
  \def\citenamefont#1{#1}\fi
\expandafter\ifx\csname url\endcsname\relax
  \def\url#1{\texttt{#1}}\fi
\expandafter\ifx\csname urlprefix\endcsname\relax\def\urlprefix{URL }\fi
\providecommand{\bibinfo}[2]{#2}
\providecommand{\eprint}[2][]{\url{#2}}

\bibitem[{\citenamefont{Matzdorf}(1998)}]{mat98}
\bibinfo{author}{\bibfnamefont{R.}~\bibnamefont{Matzdorf}},
  \bibinfo{journal}{Surf.\ Sci.\ Reps.} \textbf{\bibinfo{volume}{30}},
  \bibinfo{pages}{153} (\bibinfo{year}{1998}).

\bibitem[{\citenamefont{Schintke et~al.}(2001)\citenamefont{Schintke, Messerli,
  Pivetta, Patthey, Libioulle, Stengel, Vita, and Schneider}}]{sch01}
\bibinfo{author}{\bibfnamefont{S.}~\bibnamefont{Schintke}},
  \bibinfo{author}{\bibfnamefont{S.}~\bibnamefont{Messerli}},
  \bibinfo{author}{\bibfnamefont{M.}~\bibnamefont{Pivetta}},
  \bibinfo{author}{\bibfnamefont{F.}~\bibnamefont{Patthey}},
  \bibinfo{author}{\bibfnamefont{L.}~\bibnamefont{Libioulle}},
  \bibinfo{author}{\bibfnamefont{M.}~\bibnamefont{Stengel}},
  \bibinfo{author}{\bibfnamefont{A.~D.} \bibnamefont{Vita}}, \bibnamefont{and}
  \bibinfo{author}{\bibfnamefont{W.-D.} \bibnamefont{Schneider}},
  \bibinfo{journal}{Phys.\ Rev.\ Lett.} \textbf{\bibinfo{volume}{87}},
  \bibinfo{pages}{276801} (\bibinfo{year}{2001}).

\bibitem[{\citenamefont{Dose et~al.}(1984)\citenamefont{Dose, Altmann,
  Goldmann, Kolak, and Rogozik}}]{dos84}
\bibinfo{author}{\bibfnamefont{V.}~\bibnamefont{Dose}},
  \bibinfo{author}{\bibfnamefont{W.}~\bibnamefont{Altmann}},
  \bibinfo{author}{\bibfnamefont{A.}~\bibnamefont{Goldmann}},
  \bibinfo{author}{\bibfnamefont{U.}~\bibnamefont{Kolak}}, \bibnamefont{and}
  \bibinfo{author}{\bibfnamefont{J.}~\bibnamefont{Rogozik}},
  \bibinfo{journal}{Phys.\ Rev.\ Lett.} \textbf{\bibinfo{volume}{52}},
  \bibinfo{pages}{1919} (\bibinfo{year}{1984}).

\bibitem[{\citenamefont{Straub and Himpsel}(1984)}]{str84}
\bibinfo{author}{\bibfnamefont{D.}~\bibnamefont{Straub}} \bibnamefont{and}
  \bibinfo{author}{\bibfnamefont{F.~J.} \bibnamefont{Himpsel}},
  \bibinfo{journal}{Phys.\ Rev.\ Lett.} \textbf{\bibinfo{volume}{52}},
  \bibinfo{pages}{1922} (\bibinfo{year}{1984}).

\bibitem[{\citenamefont{Giesen et~al.}(1985)\citenamefont{Giesen, Hage,
  Himpsel, Riess, and Steinmann}}]{gie85}
\bibinfo{author}{\bibfnamefont{K.}~\bibnamefont{Giesen}},
  \bibinfo{author}{\bibfnamefont{F.}~\bibnamefont{Hage}},
  \bibinfo{author}{\bibfnamefont{F.~J.} \bibnamefont{Himpsel}},
  \bibinfo{author}{\bibfnamefont{H.~J.} \bibnamefont{Riess}}, \bibnamefont{and}
  \bibinfo{author}{\bibfnamefont{W.}~\bibnamefont{Steinmann}},
  \bibinfo{journal}{Phys.\ Rev.\ Lett.} \textbf{\bibinfo{volume}{55}},
  \bibinfo{pages}{300} (\bibinfo{year}{1985}).

\bibitem[{\citenamefont{Hertel et~al.}(1996)\citenamefont{Hertel, Knoesel,
  Wolf, and Ertl}}]{her96}
\bibinfo{author}{\bibfnamefont{T.}~\bibnamefont{Hertel}},
  \bibinfo{author}{\bibfnamefont{E.}~\bibnamefont{Knoesel}},
  \bibinfo{author}{\bibfnamefont{M.}~\bibnamefont{Wolf}}, \bibnamefont{and}
  \bibinfo{author}{\bibfnamefont{G.}~\bibnamefont{Ertl}},
  \bibinfo{journal}{Phys.\ Rev.\ Lett.} \textbf{\bibinfo{volume}{76}},
  \bibinfo{pages}{535} (\bibinfo{year}{1996}).

\bibitem[{\citenamefont{Straub and Himpsel}(1986)}]{str86}
\bibinfo{author}{\bibfnamefont{D.}~\bibnamefont{Straub}} \bibnamefont{and}
  \bibinfo{author}{\bibfnamefont{F.~J.} \bibnamefont{Himpsel}},
  \bibinfo{journal}{Phys.\ Rev.\ B} \textbf{\bibinfo{volume}{33}},
  \bibinfo{pages}{2256} (\bibinfo{year}{1986}).

\bibitem[{\citenamefont{Quiniou et~al.}(1993)\citenamefont{Quiniou, Bulvovi\'c,
  and Osgood}}]{qui93}
\bibinfo{author}{\bibfnamefont{B.}~\bibnamefont{Quiniou}},
  \bibinfo{author}{\bibfnamefont{V.}~\bibnamefont{Bulvovi\'c}},
  \bibnamefont{and} \bibinfo{author}{\bibfnamefont{R.~M.}
  \bibnamefont{Osgood}}, \bibinfo{journal}{Phys.\ Rev.\ B}
  \textbf{\bibinfo{volume}{47}}, \bibinfo{pages}{15890} (\bibinfo{year}{1993}).

\bibitem[{\citenamefont{Yang et~al.}(1993)\citenamefont{Yang, Bartynski,
  Kochanski, Papadia, Fond\'en, and Persson}}]{yan93}
\bibinfo{author}{\bibfnamefont{S.}~\bibnamefont{Yang}},
  \bibinfo{author}{\bibfnamefont{R.~A.} \bibnamefont{Bartynski}},
  \bibinfo{author}{\bibfnamefont{G.~P.} \bibnamefont{Kochanski}},
  \bibinfo{author}{\bibfnamefont{S.}~\bibnamefont{Papadia}},
  \bibinfo{author}{\bibfnamefont{T.}~\bibnamefont{Fond\'en}}, \bibnamefont{and}
  \bibinfo{author}{\bibfnamefont{M.}~\bibnamefont{Persson}},
  \bibinfo{journal}{Phys.\ Rev.\ Lett.} \textbf{\bibinfo{volume}{70}},
  \bibinfo{pages}{849} (\bibinfo{year}{1993}).

\bibitem[{\citenamefont{Petaccia et~al.}(1999)\citenamefont{Petaccia, Grill,
  Zangrando, and Modesti}}]{pet99}
\bibinfo{author}{\bibfnamefont{L.}~\bibnamefont{Petaccia}},
  \bibinfo{author}{\bibfnamefont{L.}~\bibnamefont{Grill}},
  \bibinfo{author}{\bibfnamefont{M.}~\bibnamefont{Zangrando}},
  \bibnamefont{and} \bibinfo{author}{\bibfnamefont{S.}~\bibnamefont{Modesti}},
  \bibinfo{journal}{Phys.\ Rev.\ Lett.} \textbf{\bibinfo{volume}{82}},
  \bibinfo{pages}{386} (\bibinfo{year}{1999}).

\bibitem[{\citenamefont{Echenique and Pendry}(1978)}]{ech78}
\bibinfo{author}{\bibfnamefont{P.~M.} \bibnamefont{Echenique}}
  \bibnamefont{and} \bibinfo{author}{\bibfnamefont{J.~B.}
  \bibnamefont{Pendry}}, \bibinfo{journal}{J. Phys. C: Solid State Phys.}
  \textbf{\bibinfo{volume}{11}}, \bibinfo{pages}{2065} (\bibinfo{year}{1978}).

\bibitem[{\citenamefont{Echenique et~al.}(1985)\citenamefont{Echenique, Flores,
  and Sols}}]{ech85}
\bibinfo{author}{\bibfnamefont{P.~M.} \bibnamefont{Echenique}},
  \bibinfo{author}{\bibfnamefont{F.}~\bibnamefont{Flores}}, \bibnamefont{and}
  \bibinfo{author}{\bibfnamefont{F.}~\bibnamefont{Sols}},
  \bibinfo{journal}{Phys.\ Rev.\ Lett.} \textbf{\bibinfo{volume}{55}},
  \bibinfo{pages}{2348} (\bibinfo{year}{1985}).

\bibitem[{\citenamefont{Bausells and Echenique}(1986)}]{bau86}
\bibinfo{author}{\bibfnamefont{J.}~\bibnamefont{Bausells}} \bibnamefont{and}
  \bibinfo{author}{\bibfnamefont{P.~M.} \bibnamefont{Echenique}},
  \bibinfo{journal}{Phys.\ Rev.\ B} \textbf{\bibinfo{volume}{33}},
  \bibinfo{pages}{1471} (\bibinfo{year}{1986}).

\bibitem[{\citenamefont{Chulkov et~al.}(1998)\citenamefont{Chulkov,
  Sarr\'\i{}a, Silkin, Pitarke, and Echenique}}]{chu98}
\bibinfo{author}{\bibfnamefont{E.~V.} \bibnamefont{Chulkov}},
  \bibinfo{author}{\bibfnamefont{I.}~\bibnamefont{Sarr\'\i{}a}},
  \bibinfo{author}{\bibfnamefont{V.~M.} \bibnamefont{Silkin}},
  \bibinfo{author}{\bibfnamefont{J.~M.} \bibnamefont{Pitarke}},
  \bibnamefont{and} \bibinfo{author}{\bibfnamefont{P.~M.}
  \bibnamefont{Echenique}}, \bibinfo{journal}{Phys.\ Rev.\ Lett.}
  \textbf{\bibinfo{volume}{80}}, \bibinfo{pages}{4947} (\bibinfo{year}{1998}).

\bibitem[{\citenamefont{Hedin and Lundqvist}(1969)}]{hed69}
\bibinfo{author}{\bibfnamefont{L.}~\bibnamefont{Hedin}} \bibnamefont{and}
  \bibinfo{author}{\bibfnamefont{S.}~\bibnamefont{Lundqvist}},
  \emph{\bibinfo{title}{Solid State Physics}}, vol.~\bibinfo{volume}{23}
  (\bibinfo{publisher}{Academic Press}, \bibinfo{year}{1969}).

\bibitem[{\citenamefont{Eguiluz et~al.}(1992)\citenamefont{Eguiluz,
  Heinrichsmeier, Fleszar, and Hanke}}]{egu92}
\bibinfo{author}{\bibfnamefont{A.~G.} \bibnamefont{Eguiluz}},
  \bibinfo{author}{\bibfnamefont{M.}~\bibnamefont{Heinrichsmeier}},
  \bibinfo{author}{\bibfnamefont{A.}~\bibnamefont{Fleszar}}, \bibnamefont{and}
  \bibinfo{author}{\bibfnamefont{W.}~\bibnamefont{Hanke}},
  \bibinfo{journal}{Phys.\ Rev.\ Lett.} \textbf{\bibinfo{volume}{68}},
  \bibinfo{pages}{1359} (\bibinfo{year}{1992}).

\bibitem[{\citenamefont{Deisz et~al.}(1993)\citenamefont{Deisz, Eguiluz, and
  Hanke}}]{dei93}
\bibinfo{author}{\bibfnamefont{J.~J.} \bibnamefont{Deisz}},
  \bibinfo{author}{\bibfnamefont{A.~G.} \bibnamefont{Eguiluz}},
  \bibnamefont{and} \bibinfo{author}{\bibfnamefont{W.}~\bibnamefont{Hanke}},
  \bibinfo{journal}{Phys.\ Rev.\ Lett.} \textbf{\bibinfo{volume}{71}},
  \bibinfo{pages}{2793} (\bibinfo{year}{1993}).

\bibitem[{\citenamefont{Heinrichsmeier
  et~al.}(1998)\citenamefont{Heinrichsmeier, Fleszar, Hanke, and
  Eguiluz}}]{hei98}
\bibinfo{author}{\bibfnamefont{M.}~\bibnamefont{Heinrichsmeier}},
  \bibinfo{author}{\bibfnamefont{A.}~\bibnamefont{Fleszar}},
  \bibinfo{author}{\bibfnamefont{W.}~\bibnamefont{Hanke}}, \bibnamefont{and}
  \bibinfo{author}{\bibfnamefont{A.~G.} \bibnamefont{Eguiluz}},
  \bibinfo{journal}{Phys.\ Rev.\ B} \textbf{\bibinfo{volume}{57}},
  \bibinfo{pages}{14974} (\bibinfo{year}{1998}).

\bibitem[{\citenamefont{Fratesi et~al.}(unpublished)\citenamefont{Fratesi,
  Brivio, and Molinari}}]{fra03}
\bibinfo{author}{\bibfnamefont{G.}~\bibnamefont{Fratesi}},
  \bibinfo{author}{\bibfnamefont{G.~P.} \bibnamefont{Brivio}},
  \bibnamefont{and} \bibinfo{author}{\bibfnamefont{L.~G.}
  \bibnamefont{Molinari}}, \bibinfo{journal}{cond-mat/0305344}
  (\bibinfo{year}{unpublished}).

\bibitem[{\citenamefont{Inglesfield}(1981)}]{ing81}
\bibinfo{author}{\bibfnamefont{J.~E.} \bibnamefont{Inglesfield}},
  \bibinfo{journal}{J. Phys. C: Solid State Phys.}
  \textbf{\bibinfo{volume}{14}}, \bibinfo{pages}{3795} (\bibinfo{year}{1981}).

\bibitem[{\citenamefont{Kohn and Sham}(1965)}]{koh65}
\bibinfo{author}{\bibfnamefont{W.}~\bibnamefont{Kohn}} \bibnamefont{and}
  \bibinfo{author}{\bibfnamefont{L.~J.} \bibnamefont{Sham}},
  \bibinfo{journal}{Phys.\ Rev.} \textbf{\bibinfo{volume}{140}},
  \bibinfo{pages}{A1133} (\bibinfo{year}{1965}).

\bibitem[{\citenamefont{Perdew and Zunger}(1981)}]{per81}
\bibinfo{author}{\bibfnamefont{J.~P.} \bibnamefont{Perdew}} \bibnamefont{and}
  \bibinfo{author}{\bibfnamefont{A.}~\bibnamefont{Zunger}},
  \bibinfo{journal}{Phys.\ Rev.\ B} \textbf{\bibinfo{volume}{23}},
  \bibinfo{pages}{5048} (\bibinfo{year}{1981}).

\bibitem[{\citenamefont{White et~al.}(1998)\citenamefont{White, Godby, Rieger,
  and Needs}}]{whi98}
\bibinfo{author}{\bibfnamefont{I.~D.} \bibnamefont{White}},
  \bibinfo{author}{\bibfnamefont{R.~W.} \bibnamefont{Godby}},
  \bibinfo{author}{\bibfnamefont{M.~M.} \bibnamefont{Rieger}},
  \bibnamefont{and} \bibinfo{author}{\bibfnamefont{R.~J.} \bibnamefont{Needs}},
  \bibinfo{journal}{Phys.\ Rev.\ Lett.} \textbf{\bibinfo{volume}{80}},
  \bibinfo{pages}{4265} (\bibinfo{year}{1998}).

\bibitem[{\citenamefont{Papadia et~al.}(1990)\citenamefont{Papadia, Persson,
  and Salmi}}]{pap90}
\bibinfo{author}{\bibfnamefont{S.}~\bibnamefont{Papadia}},
  \bibinfo{author}{\bibfnamefont{M.}~\bibnamefont{Persson}}, \bibnamefont{and}
  \bibinfo{author}{\bibfnamefont{L.-A.} \bibnamefont{Salmi}},
  \bibinfo{journal}{Phys.\ Rev.\ B} \textbf{\bibinfo{volume}{41}},
  \bibinfo{pages}{10237} (\bibinfo{year}{1990}).

\bibitem[{\citenamefont{Lindgren and Walld\'en}(1989)}]{lin89}
\bibinfo{author}{\bibfnamefont{S.~A.} \bibnamefont{Lindgren}} \bibnamefont{and}
  \bibinfo{author}{\bibfnamefont{L.}~\bibnamefont{Walld\'en}},
  \bibinfo{journal}{Phys.\ Rev.\ B} \textbf{\bibinfo{volume}{40}},
  \bibinfo{pages}{11546} (\bibinfo{year}{1989}).

\bibitem[{\citenamefont{Ishida}(2001)}]{ish01}
\bibinfo{author}{\bibfnamefont{H.}~\bibnamefont{Ishida}},
  \bibinfo{journal}{Phys.\ Rev.\ B} \textbf{\bibinfo{volume}{63}},
  \bibinfo{pages}{165409} (\bibinfo{year}{2001}).

\bibitem[{\citenamefont{Bulovi\'c et~al.}(1994)\citenamefont{Bulovi\'c,
  Quiniou, and Osgood}}]{bul94}
\bibinfo{author}{\bibfnamefont{V.}~\bibnamefont{Bulovi\'c}},
  \bibinfo{author}{\bibfnamefont{B.}~\bibnamefont{Quiniou}}, \bibnamefont{and}
  \bibinfo{author}{\bibfnamefont{R.~M.} \bibnamefont{Osgood}},
  \bibinfo{journal}{J.\ Vac.\ Sci.\ Technol.\ A} \textbf{\bibinfo{volume}{12}},
  \bibinfo{pages}{2201} (\bibinfo{year}{1994}).

\end{thebibliography}

\end{document}